# A Model-driven Approach for Grid Services Engineering

**David Manset [1,2,3], Hervé Verjus [3], Richard McClatchey [1], Flavio Oquendo [4]**

[1] CCCS, Centre for Complex Co-operative Systems, University West of England, Bristol BS16 1QY, UK

[2] DSU-TT Division, CERN 1211, Geneva 23, Switzerland

[3] LISTIC, University of Savoie, Annecy, France

[4] VALORIA, University of South Brittany, Vannes, France

**Abstract**

As a consequence to the hype of Grid computing, such systems have seldom been designed using formal techniques. The complexity and rapidly growing demand around Grid technologies has favour the use of classical development techniques, resulting in no guidelines or rules and unstructured engineering processes. This paper advocates a formal approach to Grid applications development in an effort to contribute to the rigorous development of Grids software architectures. This approach addresses cross-platform interoperability and quality of service; the model-driven paradigm is applied to a formal architecture-centric engineering method in order to benefit from the formal semantic description power in addition to model-based transformations. The result of such a novel combined concept promotes the re-use of design models and eases developments in Grid computing by providing an adapted development process and ensuring correctness at each design level.

# 1. INTRODUCTION

The new Grid paradigm has been described as "*a distributed computing infrastructure for advance science and engineering*" that can address the concept of *"coordinated resource sharing and problem solving in dynamic, multi-institutional virtual organizations"* in [1, 2, 3]. This coordinated sharing is not only file exchange but can also provide direct access to computers, software, data and other system resources. Grid applications bundle different services using a heterogeneous pool of resources in a so-called *virtual organization*. This makes Grid applications very difficult to model and to implement. In addition, one of the major issues in today's Grid engineering is that it often follows a code-driven approach. As a direct consequence the resulting source code is neither re-usable nor does it promote dynamic adaptation facilities as it should, since it is a representation of the Service Oriented Architecture (SOA) paradigm [4, 5]. As it is directly extracted from the definition of the Grid concept, Grid design should ensure cross-platform interoperability by providing ways to re-use concretely systems in a heterogeneous context. Having no guidelines or rules in the design of a Grid-based application is a paradox since there are many existing architectural approaches for distributed computing applications which could ease the engineering process, could enable rigorous engineering methods and could promote the re-use [6] of components in future Grid developments. Although it has been proven from past experience that using structured engineering methods would ease the development process of any computing system and would reduce the complexity when building large Grid applications, the hype of Grid computing has been and still is forcing brute force coding and rather unstructured engineering processes. This always leads to a loss of performance, interoperability problems and generally results in very complex systems that only dedicated developers can manage.

It is our belief that semi-formal engineering methods in current use are insufficient to tackle tomorrow's requirements in Grid computing. This paper introduces a novel approach by applying the model-driven engineering philosophy inside a formal engineering approach. Inside a well-defined and adapted formal approach, we investigate the enactment of our model-driven engineering process to complete our design framework and provide tools to build the next generation of Grid applications. The remainder of this paper emphasizes different aspects which are, in our view, essential to Grid engineering:

- offering a user-friendly vision to Grid architects by providing re-usable conceptual building blocks,
- hiding the complexity of the final execution platform through abstraction models, and finally
- promoting design re-use to ease further developments.

In order to achieve these objectives, we combine two approaches together and seek advantages from both of them. On the one hand, we use a formal semantic descriptive power to model-check Grid software architectures; on the other hand, we use a model-driven approach to promote model re-use, to hide the platform complexity and to translate abstract software descriptions to concrete usable ones.

The remainder of this paper is structured as follows. Part 2 presents the different approaches we are using (i.e. model-driven engineering (MDE) and software architecture centric approach) and then conclude on a design approach, which is a combination of these two. Part 3 explains how model-driven engineering is enacted to design Grid applications. Part 4 presents our formal architecture-centric model-driven approach and the means used to achieve it. Finally, we conclude with identifying future work to be done with respect to our framework and state the benefits of using it.

# 2. MODEL DRIVEN ENGINEERING

Model Driven Engineering (MDE) [7], probably derived from the OMG's Model Driven Architecture (MDA) initiative [8], tackles the problem of system development by promoting the usage of models as the primary artifact to be constructed and maintained. Some of the model driven engineering approaches provide languages and tools to transform models by means of transformation rules in order to describe and design complex systems at a high level of abstraction. The main objective of such methods is to hide the complexity and constraints induced by the target execution platform, during the design phase. Thus, an architect can mainly focus on functional requirements of his application rather than on non-functional ones. The next section discusses the architecture-centric paradigm, and then introduces our dedicated approach to Grid applications.

## 2-1. Architecture-centric Engineering Approach

The architecture-centric approach focuses on the software architecture description used to organise development activities. Thus every stage of the software life cycle – including specification, implementation and also architectural style [9, 10, 11, 12] – are considered as part of the process. The work on architecture-centric approaches for software development has been very fruitful during the past ten years, leading, amongst other results, to the proposition of a wide variety of Architecture Description Languages (ADLs) [13, 14, 15], usually accompanied by analysis tools. These languages are used to formalize the architecture description as well as its refinement. The benefits of using such an approach are various. They rank from the increase in architecture comprehension among the persons involved in a project (due to the use of an unambiguous language), to the re-use at the design phase and to the property description and analysis (properties of the future system can be specified and the architecture analyzed). Figure 1 introduces the architecture-centric development process provided in [16]. The enthusiasm around the development of formal languages for software architecture descriptions comes from the fact that such formalisms are suitable for automated handling and pre-implementation property checking. As discussed later in this paper, most of these aspects are essential to the enactment of the model-driven paradigm. In our approach we use this architecture-centric vision as a strong basis to further investigations in enabling MDE. The formal dimension, in addition to adapted tools, gives our MDE approach its robustness.

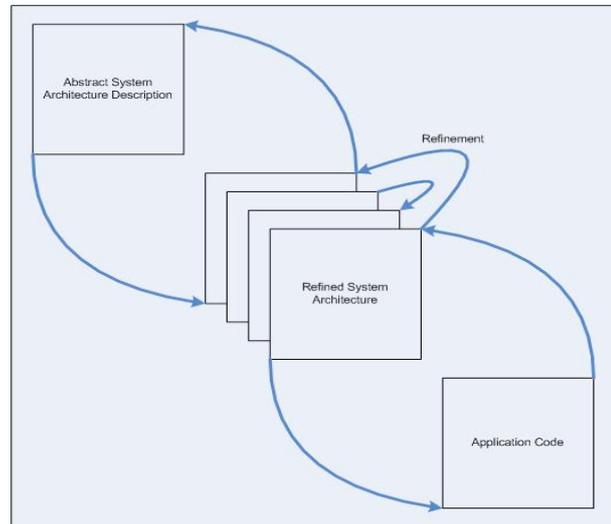

*Figure 1: Architecture-centric Development Process*

## 2-2. A combination of approaches

Focusing on the model transformation aspects, we can notice similarities with the refinement concepts found in formal architecture-centric software engineering. Although the main purposes of both concepts being slightly different (the refinement of an abstract software architecture aiming at making it more and more concrete), the MDE can benefit from refinement to handle some types of transformations. As a matter of fact, given the hypothesis that models of the system are expressed in a well formed and standard architectural language capable of refinement, it is potentially possible to apply refinement actions on models to adapt them with respect to platform constraints. Thus investigating different platforms can lead to the creation of transformation and constraint models applicable to an abstract system model. This is the reason why formal architecture-centric software engineering concepts are very well suited to the enactment of the MDE process.

Thus the combination of architecture-centric and model-driven paradigms can provide parts of the model transformations and the semantic of description to MDE. Combining these two paradigms makes them more complete with respect to the Grid application domain. This combination is explained in the next section.

## 3. A FORMAL ARCHITECTURE-CENTRIC MDE

Following our MDE paradigm, we address the challenge of designing, optimizing and refining a Grid abstract architecture, with respect to different criteria; the final objective being the automatic generation of a complete set of Grid services. From the study we conducted in Grid engineering we consider the Grid as a SOA and define a set of Quality of Services (QoS) properties [17]. Using a formal approach to describe these aspects we build a set of models and investigate the feasibility of enacting this model-driven process. The remainder of this paper is based on a Grid domain-specific vocabulary.

## 3-1. Defining the key models

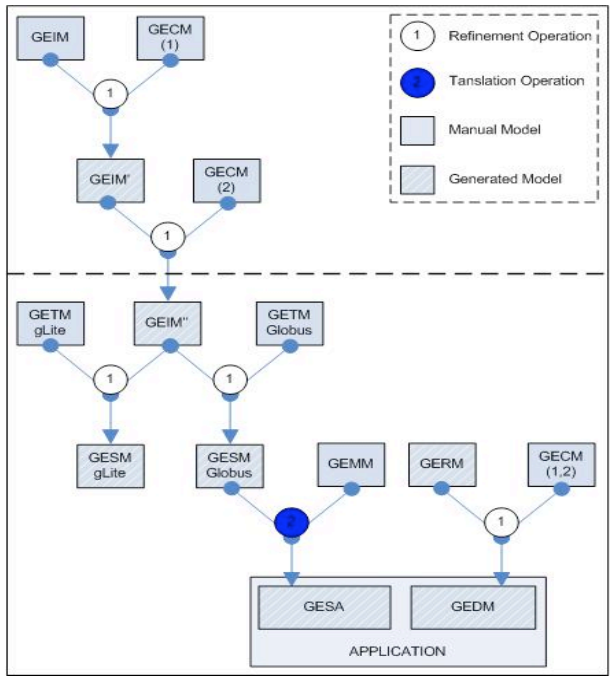

*Figure 2: Grid Model-driven Engineering Process*

In Grid engineering, design is largely effected by many constraints; these constraints are of different types and are introduced either by the architect himself when specifying properties like quality of service or by the target execution platform. Thus the MDE process dedicated to Grid engineering must take into account all of these aspects in providing the necessary models and their respecting relationships (see figure 2). By proposing several models, our approach separates concerns and addresses different aspects of the application. Thus expertise management and capture is better than in classical approaches [18, 19]. Each model represents an accurate view of the system useful for conceptual understanding and analysis. Unlike the software engineering refinement process where the architect iteratively refines the system architecture, most of the model transformations in our MDE are automated. The different models composing our process are defined as follows:

- *GEIM* – the Grid Environment Independent Model is an abstract architecture of the Grid application based on a formal ADL using specific constructs.
- *GESM* – the Grid Environment Specific Model is a concrete system architecture close to final code and optimized according to a selected execution platform (a refined system description),
- *GECM* – the Grid Execution Constraint Model is an abstract description of adaptable design patterns suitable to some QoS properties,
- *GETM* – the Grid Environment Transformation Model is an abstract description of adaptable design patterns representing a target platform.

In order to simplify our approach, we will not discuss in detail the other models; however as a clarification of the concept these models can be defined as follows:
- GEMM – the Grid Environment Mapping Model is a model of translation between an abstract description and an implementation language (i.e. in charge of defining the mapping between the semantics of our meta-model and a given programming language, for instance Java $^{TM}$).
- GERM – the Grid Environment Resource Model is a model representing the physical constitution of the Grid.
- GEDM – the Grid Environment Deployment Model is a model specifying how to deploy the resulting application onto the grid set of resources,
- GESA – the Grid Environment Specific Application is the auto-generated source code of the application (i.e. obtained after GEMM translation).

*Figure 2* introduces the orchestration of these models according to an MDE process. The upper part of the figure represents re-usable design models, whereas the lower parts represents specific models and implementations. We make a distinction between two major levels, one is the architecture level of transformation and the other is the implementation level of transformation.

In this process, two sets of QoS constraints have been introduced during design (*GECM 1 & 2*). In addition two different target platforms were also introduced to show the interest of such an approach (*GETM gLite* [20] *& Globus* [21]). The software architect defines some of these models, such as the GEIM and the GECM, unlike the others, which are automatically built from the transformation of previous models. As mentioned above, our model-driven approach uses the architecture-centric refinement concept to handle model transformations. Automating these transformations helps in decoupling architect's functional specifications and platform non-functional requirements. The next section details these transformations in terms of nature and objectives, whilst introducing the whole process.

### 3-2. The Architecture-Centric MDE Process

As explained in the previous section, our approach enacts a set of models. These models can be of two distinct types; either the model is *manually* created or it is *automatically* obtained by transformation. The transformation itself can then be of two different types too; either the transformation is a composition of one or more refinement actions to adapt the current architecture or it is a translation mapping to obtain the concrete application code.

## 4. ENACTING MDE, A CONCRETE FRAMEWORK

### 4-1. ArchWare: the Formal Architecture-centric Approach and Toolkit

ArchWare [22] is an architecture-centric engineering environment allowing design from formal descriptions of software. Such a formal method enables the support of critical correctness requirements and provides tools to guarantee system properties and reliable execution. ArchWare delivers a set of formal languages and corresponding tools to enable reliable design and refinement.

### 4-2. The ArchWare Refinement Concept

Complex systems cannot be designed in one single step. In a stepwise architecture refinement, a sequence of steps starting from an abstract model of the architecture leads to a concrete, implementation-centered model of the architecture. These refinement steps can be carried out along two directions: "vertical" and "horizontal". The concrete architecture of a large software system is often developed through a "vertical" hierarchy of related architectures. An architecture hierarchy is a linear sequence of two or more architectures that may differ with respect to a variety of aspects. For instance, an abstract architecture containing functional components related by data flow connections may be implemented in a concrete architecture in terms of procedures, control connections, and shared variables. In general, an abstract architecture is simpler and easier to understand; a concrete architecture reflects more implementation concerns. "Vertical" refinement steps add more and more detail to abstract models until the concrete architectural model has been described. A refinement step typically leads to a more detailed architectural model that increases the determinism while implying properties of the abstract model. "Horizontal" refinement concerns the application of different refinement actions at different parts of the same abstract architecture, for instance, by partitioning an abstract component into different parts at the same abstraction level. An architecture concrete model can be thought of as just another architectural model in a style suitable for implementation.

As it is mentioned in [16], the ArchWare refinement approach handles an exhaustive set of refinement actions. Architecture refinement can be carried out in a series of steps, a basic step being defined in terms of basic refinement operation that can transform an architecture. An architectural model can be refined into another more concrete (i.e. more refined) architectural model. A refinement step can be carried out by the application of one or many refinement action(s). The ArchWare ARL [23] language is the formal expression of these transformations, which aims at preserving upper abstract architecture properties while modifying it. What makes the ArchWare project original is the facility to ensure that decompositions preserve any rigorously defined properties of the parent. In terms of semantic, a refinement operation is expressed as follows:

```
on a : architecture action actionRef is refinement (
        t : type ) {
            pre is { a::types includes? t }
            post is { a::types excludes? t }
        } as { a::types excludes t }
```

As discussed in *figure 2*, most of the model transformations in our model-driven process are of the same nature: architectural refinement. As we detail it in the next section, respecting constraints introduced by platforms and QoS is a matter of refinement actions application.

### 4-3. A Refinement Process in an MDE perspective

In our MDE approach, we focus on both directions of refinement – i.e. the "vertical" and the "horizontal" as discussed in *section 4-2*. Our intention is to not only refine an architecture to a concrete and "close to final" code form, but also to optimize it according to constraints. We propose two ways of using the model-driven process in Grid engineering. The first consists of optimizing a given abstract architecture according to expressed users' requirements in terms of quality of services (QoS). The second consists of optimizing an architecture according to the target execution environment. Respectively:
- *QoS properties*: each QoS property owns its equivalent design pattern. This pattern is then applied to the current software architecture through refinement actions.
- *Platform properties*: each platform owns its representing design pattern and properties. The software architecture is then adapted with respect to the platform representation through refinement actions too.

In that context, enabling MDE requires the expression and consideration of external models of transformations by means of a semantic; in our case: ARL. As an example, the following is a simplified description of the fault-tolerance QoS property design pattern to apply onto an architectural element "b":

```
FT is qualityOfServiceProperty {
  on a:architecture actions {
     include FTConnector is connector {
        … connector description …
     }
     on b:architecturalElement actions {
        replicate b .
        unify b::port::connection with FTConnector..
        … etc …
     }
  }
}…
```

Following the same scheme, we can adapt the grid application architecture to different constraints. As an instance, we can handle performance, cost, and load-balancing constraints by applying such patterns.

Given the flexibility of our formal model-driven approach and relying on the correctness of our models, the resulting tool is able to tackle every aspect of software architectures transformations needed in Grid development. These models and their enacted process constitute the core of our grid model driven engineering environment (*gMDEnv)*.

## 5. FUTURE WORK

In this paper we presented a technique for specifying Grid applications by modeling and by transforming these models to automate the adaptation to specific platforms. Since the development of this approach is nearing completion, we are now focusing on QoS properties and their corresponding design models. We have started defining an extension to the ARL language in order to increase its descriptive power while not modifying its core semantics. This work is done in collaboration with Web Services practitioners, to make it re-usable in the context of Service composition. As a matter of fact, QoS properties defined at the design level can be useful when achieving Service compositions. In the previous sections, we described a study that illustrates how our approach can tackle QoS specifications in addition to platform requirements. Our study leads to an investigation of the most frequently used platforms in Grid computing, which will result in the required *GETM* models. The power of our approach depends mainly on the correctness of these models; consequently great care is being taken to ensure this. As a proof of concept, the engineering framework being developed is enacting the combination of the formal architecture-centric and model-driven approaches introduced previously. In its current state, it is already capable of handling most of the presented models and transformations. In future we shall investigate case studies to validate its usability and promote its user-friendliness. Since this approach is based on the concepts of re-use and execution platform independence our engineering framework is not limited to the Grid domain. The same approach could tackle other developments based on the Service Oriented Architecture vision such as web services based applications (i.e. online traders, booking systems, video on demand system etc).

Thus, the benefits of using our approach are numerous. Application models designed using our framework are persistent and re-usable, as long as they are independent of the platform. For instance, one can use libraries and previously stored models to design new applications. The approach is scalable; one could extend the scope limitation of the framework by simply providing corresponding platform constraint models. And finally, from establishment of well-known architectural concepts, the framework brings the user to a high level of description while promoting user-friendliness through a simple semi-automated graphical user interface.

Finally, with respect to model transformations, another interesting area of future research is the development of a decision support system to help users through model-driven transformations. Indeed, some of the adaptations needed to satisfy platforms can lead to critical decisions. We are using the example described previously and others, to elicit the framework requirements.

## 6. CONCLUSION

In this paper we investigated model-driven process enactment using a formal architecture-centric approach to designing Grid systems. We analyzed the needs for this paradigm and shown clearly the feasibility of its implementation using the ArchWare tools. Our method was also applied to more elaborate models specific to the Grid domain in order to demonstrate that MDE can be used from design to deployment of an application. In this vision, our model-driven approach covers all the aspects required in the development of complex distributed systems such as Grids. The approach described here extends the OMG's vision by concentrating on the detail of models and transformations; and on categorizing them into different types. This paper is a first investigation of the model-driven paradigm enactment using reliable, established formal architecture-centric concepts.

Besides supporting the usefulness of ArchWare ARL and related tools, we are able to draw a number of general conclusions. We learned that the model-driven approach is a very useful paradigm when addressing cross-platform developments and problems of re-use but it must be dependent on a rigorous basis to be efficient. The formal dimension brought by ArchWare is one of the key points of our successful implementation. Similarly we learned that QoS attributes are not easy to quantify in models. There is a true lack of standards that could help significantly when considering resource comparisons. In the context of other engineering frameworks and given the basic concepts we have now in hand, our approach can provide directly relevant benefits to the practice of Grid system engineering. From our experience, we believe that MDE approach is an important contribution to the development of new Grid systems.

## ACKNOWLEDGEMENTS

The authors wish to thank their Home Institutions and the European Commission for financial support in the current research and to gratefully acknowledge Karim Megzari for his contribution on the refinement aspects of software architectures.